\documentclass[pra,twocolumn,superscriptaddress]{revtex4}
\usepackage{blindtext}
\usepackage{amsfonts}
\usepackage{amsmath,amssymb}

\usepackage{graphicx}
\usepackage{subfigure}
\usepackage{bm}
\usepackage{epstopdf}
\begin{document}
\title{Quantum Simulation of Competing Orders with Fermions in  Quantum Optical Lattices}

\author{Arturo Camacho-Guardian } 
\affiliation{ 
 Instituto de F\'{\i}sica, Universidad
Nacional Aut\'onoma de M\'exico, Apartado Postal 20-364, M\'exico D.
F. 01000, Mexico. }
 \affiliation{ 
Department of Physics and Astronomy, Aarhus University, Ny Munkegade, DK-8000 Aarhus C, Denmark. }           
\author{Rosario Paredes} 
\affiliation{ 
 Instituto de F\'{\i}sica, Universidad
Nacional Aut\'onoma de M\'exico, Apartado Postal 20-364, M\'exico D.
F. 01000, Mexico. }
 \author{Santiago F. Caballero-Ben\'itez}
 \email{scaballero@fisica.unam.mx}
 \affiliation{ 
 Instituto de F\'{\i}sica, Universidad
Nacional Aut\'onoma de M\'exico, Apartado Postal 20-364, M\'exico D.
F. 01000, Mexico. }       
\affiliation{CONACYT-Instituto Nacional de Astrof\'{\i}sica, \'Optica y Electr\'onica, Calle Luis Enrique Erro No. 1, Sta. Mar\'{\i}a Tonantzintla, Pue. CP 72840, M\'exico
}

%\pacs{67.85.-d, 03.75.Ss, 03.75.Gg}

\begin{abstract}
Ultracold Fermi atoms confined in optical lattices coupled to quantized modes of an optical cavity are an ideal scenario to engineer quantum simulators in the~strongly interacting regime. The~system has both short range and cavity induced long range interactions. We propose such a scheme to investigate the~coexistence of superfluid pairing, density order and quantum domains having antiferromagnetic or density order in the Hubbard model in a high finesse optical cavity at $T=0$. We demonstrate that those phases can be accessed by properly tuning the linear polarizer of an external pump beam via the cavity back-action effect, while modulating the system doping. This allows emulate the typical scenarios of analog strongly correlated electronic systems.  
\end{abstract}

\maketitle

%\section{Introduction}
\textit{Introduction.} Coupling ultracold quantum gases to high-finesse optical cavities is a novel scenario to explore many-body phases in the full quantum regime by exploiting the controllability of light-matter interaction~\cite{Brennecke, Mekhov1}. Major experimental breakthroughs have been achieved in the quantum limit of both light and matter. For instance, the Dicke phase transition has been observed in a Bose-Einstein condensate coupled to cavity modes~\cite{Klinder}. Experimentally, it has been achieved the emergence and control of supersolid phases where the cavity backaction  generates light-induced effective long-range interactions which compete with short-range interatomic interactions~\cite{Leonard, Landig,EsslingerHiggs,Ketterle,Hruby}. On the theoretical side, recent studies have introduced settings where cavity fields generate gauge-fields~\cite{Kollath,Ballantine}, artificial spin-orbit coupling~\cite{Deng}, self-organized phases \cite{Gopalakrishnan,Ritsch}, topological phases~\cite{Pan,Mivehvar}, measurement induced entangled modes~\cite{PRLElliot}, induced magnetic and density order using  measurement back action~\cite{Gabriel2} and feedback control~\cite{Gabriel3}, dimerization~\cite{Caballero4}, spin lattice systems~\cite{Morigi} and quantum simulators based on global collective light-matter interactions~\cite{Santiago1, Santiago3}. 

Ultracold Fermi gases loaded in optical lattices (OL) with long-range interactions using magnetic polar interactions and electric dipolar interactions are possible~\cite{Gadway}. However, the temperatures needed to investigate the competition between different orders and the fact that interactions depend on particular constituents pose some limitations. Additionally, extended Bose-Hubbard systems~\cite{Ferlaino} have been achieved, Rydberg systems~\cite{Gelhausen} have been proposed, while small systems with ions~\cite{ions} or analogous superconducting circuits~\cite{Leek} and multi-mode opto-mechanics are also possible~\cite{optomechanics}.  Ultracold gases in OL inside cavities allow engineering of spatial structure in many-body interactions, that beyond dipolar systems, is independent of the interaction intrinsic nature. The effective light-induced many-body interaction can be engineered externally, controlling the properties of the light pumped into the system.
\begin{figure}[!t]
\begin{center}
\includegraphics[width=3.2in]{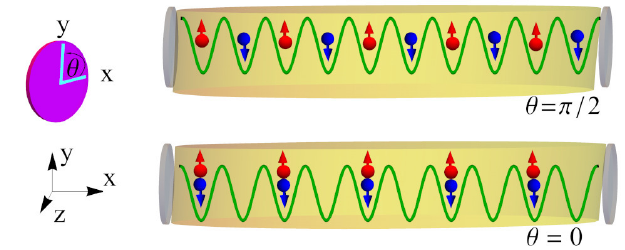} 
\end{center}
\caption{\footnotesize(Color online) Schematic representation of the model. A Fermi-Hubbard chain is placed inside a single mode standing wave optical cavity.  A linearly polarized pump beam with incidence perpendicular to the cavity along the $z$ axis, creates an electromagnetic field that couples light to the atomic modes. The polarization disk (purple) shows the choice of polarization angle $\theta$ measured from the -$x$ polarization axis. The light pumped into the system induces density wave order maximally for $\theta=0$ (bottom) while antiferromagnetic order is maximal for $\theta=\pi/2$ (top).}
\label{Fig1}
\end{figure}

Simulating quantum many-body long-range Hamiltonians with emergent quantum phases of matter complements current efforts to understand condensed matter phenomena~\cite{Lewenstein}. In contrast to condensed matter systems, ultracold quantum gases can be measured with a single site resolution~\cite{Parsons,Boll,Cheuk,Mazurenko}. This allows extracting of density distributions and non-local correlation functions~\cite{Endres}. Using this information in experiments, it is possible to produce and detect the properties of 1D Hubbard chains seeking for hidden antiferromagnetic correlations~\cite{Hilker} with the aim of improving our knowledge on the phase diagram of the Hubbard model~\cite{Cocchi, Greif, Hart}.

In this article, we propose a route to study the emergence of quantum phases of fermionic matter resulting from the~competition between short and long-range light-induced interactions via cavity fields. The latter type of interactions is absent in standard (``classical") optical lattices~\cite{Bloch,Lewenstein} without high-Q cavities. For this purpose, we consider a mixture of Fermi atoms in two hyperfine states confined in an optical lattice in one dimension.  The atomic chain lies in a single mode cavity, illuminated by a coherent pump beam, see Fig.\ref{Fig1}. We tune the polarization angle of the pump beam and show that this allows to manipulate the emergence of different phases of quantum matter. The system will support the formation of density wave (DW) insulators, antiferromagnets~\cite{Mazzucchi}, pair-superfluid states (SF$_\eta$)~\cite{Yang} or pair density waves (PDW)~\cite{Fradkin, Hamidian}.  We show that by controlling the polarization of the pump beam, it is possible to break different system symmetries. Therefore, we can instigate the formation of the aforementioned phases at will, with full parametric control by external means using the pump beam. We found that quantum domains are formed with antiferromagnetic or density wave character depending on the hole/pair doping in the system.

\textit{Model.} The system consists of a mixture of Fermi atoms in two hyperfine spin states confined in an optical lattice in one dimension. The lattice, placed along ${\text x}$ direction, is inside of a single mode cavity with frequency $\omega_c$ far from the atomic resonance $\omega_a$. The atoms couple to the cavity via the effective coupling coefficients $g_{p, \sigma}$ for each spin polarization labeled by $\sigma$, and the atomic detuning is given by $\Delta_a=\omega_p-\omega_a$. The cavity is characterized by a decay rate $\kappa$, this condition implies that the atomic spontaneous emission of the atoms is much smaller than the detuning among the pumping mode of light $\omega_p$ and the atomic frequency $\omega_a$. The total Hamiltonian of the system in the pump reference frame is given by\cite{Caballero4,Santiago3,NJPhys2015},
\begin{equation}
\hat{\mathcal{H}}=\hat{\mathcal{H}}^a+\hat{\mathcal{H}}^f+\hat{\mathcal{H}}^{af}
\end{equation}
with the atomic dynamics contained in $\hat{\mathcal{H}}^f$, the part corresponding to the photons in the cavity described by $\hat{\mathcal{H}}^a$ and the interaction between the photons in the system and the atoms mediated by the cavity included in $\hat{\mathcal{H}}^{af}$.
The atomic dynamics is governed by  the standard Hubbard Hamiltonian
 \begin{equation}
 \hat{\mathcal{H}}^f=-t\sum_{\sigma}\sum_{\langle i , j \rangle }\left(\hat{f}_{i\sigma}^\dagger \hat f_{i\sigma}^{\phantom{\dagger}}+\textrm{H.c.}\right)+U\sum_{i}\hat{n}_{i\uparrow}\hat{n}_{i,\downarrow},
 \end{equation}
 where $\hat{f}_{i,\sigma}^{\phantom{\dagger}},\hat{f}^\dagger_{i\sigma}$ denote the annihilation and creation operators of fermions with spin $\sigma$ at site $i$ and $\hat{n}_{i\sigma}=\hat{f}^{\dagger}_{i,\sigma}\hat{f}_{i\sigma}^{\phantom{\dagger}}$ the corresponding number operator, $U$ is the interspecies on-site interaction strength, and $t$ is the tunnelling amplitude in the single band approximation. We consider lattice depths where the single band approximation is valid, namely, $V_0\gtrsim5E_R$, being $E_R$ the recoil energy.  The light part of the Hamiltonian is $\hat{\mathcal{H}}^a=-\Delta_c\hat a^\dagger\hat a$, being $\Delta_c=\omega_p-\omega_c$ the cavity-pump detuning. The dispersive shift is also included in this definition. The light-matter interaction coupled by the cavity is given by 
 \begin{equation}
\hat{\mathcal{H}}^{af}=\sum_{\sigma}(g^*_{p,\sigma}\hat{a}_c\hat{F}_{\sigma}^{\dagger}+g_{p,\sigma}\hat{a}^\dagger_{c}\hat{F}_{\sigma}^{\phantom{\dagger}}),
\label{eqaf}
 \end{equation}
 where $g_{p,\sigma}=g_c g_{p} a_{p,\sigma}/\Delta_{a}$ is the effective two photon Rabi frequency for each spin polarization, $a_{p,\sigma}$ are the coherent pump amplitudes for each spin polarization, with $g_p$ and $g_c$ the light-matter coupling coefficients of the cavity and the pump modes. The spatial projection of pump and cavity mode onto the atoms is given by \cite{Gabriel2},
 \begin{equation}
 \hat{F}_\sigma=\int d\,\vec{r}\, u_{c}^*(\vec{r})u_{p}(\vec{r})\hat{n}_{\sigma}(\vec{r}),
 \end{equation}
 with $ u_{c,p}^*(\vec{r})$ the cavity and pump mode functions, and $\hat{n}_{\sigma}(\vec{r})$ the density field of the atoms for each spin projection. 
 
It is convenient  to expand the atomic fields in terms of the Wannier functions $w(\vec{r}-\vec{r}_j)$~\cite{Mekhov1},
   \begin{equation}
   J_{ij}=\int d\vec{r}\, w(\vec{r}-\vec{r}_i)u_{c}^*(\vec{r})u_p(\vec{r})w(\vec{r}-\vec{r}_j).
   \end{equation} 
  Assuming well-localized atoms ($V_0\sim 10E_R$), we can neglect off-diagonal coupling contributions given by the inter-site overlap integrals, $i\neq j$. Off-diagonal coupling terms may lead to exotic states of matter and bond ordered states as in~the bosonic analog~\cite{Santiago3,Caballero4} and will be considered elsewhere. With this,  $\hat{F}_\sigma\approx \hat{D}_\sigma$, where $\hat{D}_\sigma=\sum_i J_{ii} \hat{n}_{i,\sigma}$ is the diagonal coupling of light to on-site atomic densities. In a one dimensional optical lattice and considering standing waves as mode functions for the light modes, we have $u_{c}(\vec{r})=\cos(\vec{k}_c\cdot\vec{r}+\phi_c)$ and $u_{p}(\vec{r})=\cos(\vec{k}_p\cdot\vec{r}+\phi_p)$. Thus, the structure of the $J_{ii}$ coefficient can be controlled by properly selecting the orientation of the cavity with respect to the optical lattice and the angle of incidence of the pump beam. We consider $\vec{k}_p=k_p\hat{e}_z$, and $\vec{k}_c=\pi/a\hat{e}_x$ ($\lambda=2a $), with $k_p=0$ for simplicity~\cite{PRAWojciech2015}. The pump and the cavity are at $90^\circ$ with respect to each other and the cavity is in the same plane as the classical optical lattice, along $\hat{e}_x$. 
 Therefore, we have 
  $\hat{D}_\sigma=J_D\sum_{i}(-1)^i \hat{n}_{i,\sigma}$,
 with $J_D=\mathcal{F}[W_0]\left(\frac{\pi}{a}\right)\cos(\phi_p)\cos(\phi_c)$, 
 where $\mathcal{F}[W_0](\vec{k})$ is the Fourier transform of $W_0(\vec{r})=w^2(\vec{r})$. The phases $\phi_{c,p}$ can be chosen arbitrarily, for simplicity we consider $\phi_{c,p}=0$. For a classical optical lattice with depth $V_0\sim10E_R$, typically  $J_D\sim1$, 
   
\textit{Light polarization and cavity back-action}. The operators concerning the light modes are given by  $\hat{a}_c$ for the cavity mode while $a_{p,R/L}$ is a classical coherent pump describing the light polarization in the circular polarization basis ($L$ or $R$). The pump mode in the ($L$, $R$) basis for an arbitrary orientation $\theta$ can be written as
  \begin{equation} 
  a_{p,\theta}=\frac{1}{\sqrt{2}}(a_{p,L}e^{i\theta}+a_{p,R}e^{-i\theta}),
  \end{equation}
  with $a_{p,R/L}\neq0$. Note that $a_{p,\theta=0}$ is the linear polarization in -$x$ while $ a_{p,\theta=\pi/2}$ corresponds to -$y$ polarization. From the Hamiltonian, the steady state of light given by the stationary limit of the Heisenberg equation of motion for the light field leads to \begin{equation}
  \hat{a}_c\approx\sum_{\sigma}C_\sigma \hat D_{\sigma},\label{eqes}
  \end{equation} 
  where $C_{\sigma}=g_{p,\sigma}/(\Delta_{c}+i\kappa)$ and  $\kappa$ has been phenomenologically introduced \cite{Mekhov1}. Equivalently, using $L$ and $R$ polarization basis, we can write $\hat{a}_c=C_\theta \hat D_\theta$, where,
    \begin{equation}
\hat{D}_\theta=\sum_{j} (-1)^j(\hat{n}_{j,\uparrow}e^{i\theta}+\hat{n}_{j,\downarrow}e^{-i\theta}),
\end{equation}  
with the polarization angle $\theta$,  $C_\theta=\tilde g_{p}/(\Delta_{c}+i\kappa)$ and $\tilde g_{p}=g_c g_p a_{p,\theta}/\Delta_a$. Which is a valid parametrization  except from the points where $a_{p,\uparrow}=0$ or $a_{p,\downarrow}=0$, having only one of the spin polarizations coupled. The pump beam intensity $|a_{p,\theta}|^2$ is  fixed typically in experiments. 
  
\textit{Effective Hamiltonian.} In the limit where the cavity back-action is dominant and measurement back-action is negligible ($\Delta_c\gg\kappa$) in the steady state of light \cite{NJPhys2015}, the effective matter Hamiltonian can be obtained by using the steady state solution\eqref{eqes} in the adiabatic limit and the full light-matter Hamiltonian equation \eqref{eqaf}. The effective matter Hamiltonian is given by
\begin{equation} 
\hat{\mathcal{H}}_{\mathrm{eff}}=\hat{\mathcal{H}}^f+\frac{g_{\mathrm{eff}}}{2 N_s}(\hat{D}^\dagger_\theta \hat{D}_\theta^{\phantom{\dagger}}+\hat{D}_\theta^{\phantom{\dagger}} \hat{D}^\dagger_\theta),
\label{eqf}
\end{equation}
where the effective coupling strength is given by
\begin{equation}
g_{\mathrm{eff}}=\frac{\Delta_{c}|\tilde g_{p}|^2|J_D|^2N_s }{\Delta_{c}^2+\kappa^2},
\end{equation}
The effective light-induced structured infinite range interaction energy is described by the last term of the Hamiltonian. This is the quantum optical lattice (QOL) contribution~\cite{Mekhov1,Santiago3} that depends on the cavity back-action of the system. We refer to the effective Hamiltonian describing the system as a fermionic quantum optical lattice Hamiltonian (FQOL). The pump-cavity detuning allows controlling the sign of $g_{\mathrm{eff}}\approx |\tilde g_p|^2|J_D|^2N_s/\Delta_{c}\propto V_0/\Delta_c$ \cite{Landig}.

 \begin{figure}[!t]
\begin{center}
\includegraphics[width=3.4in]{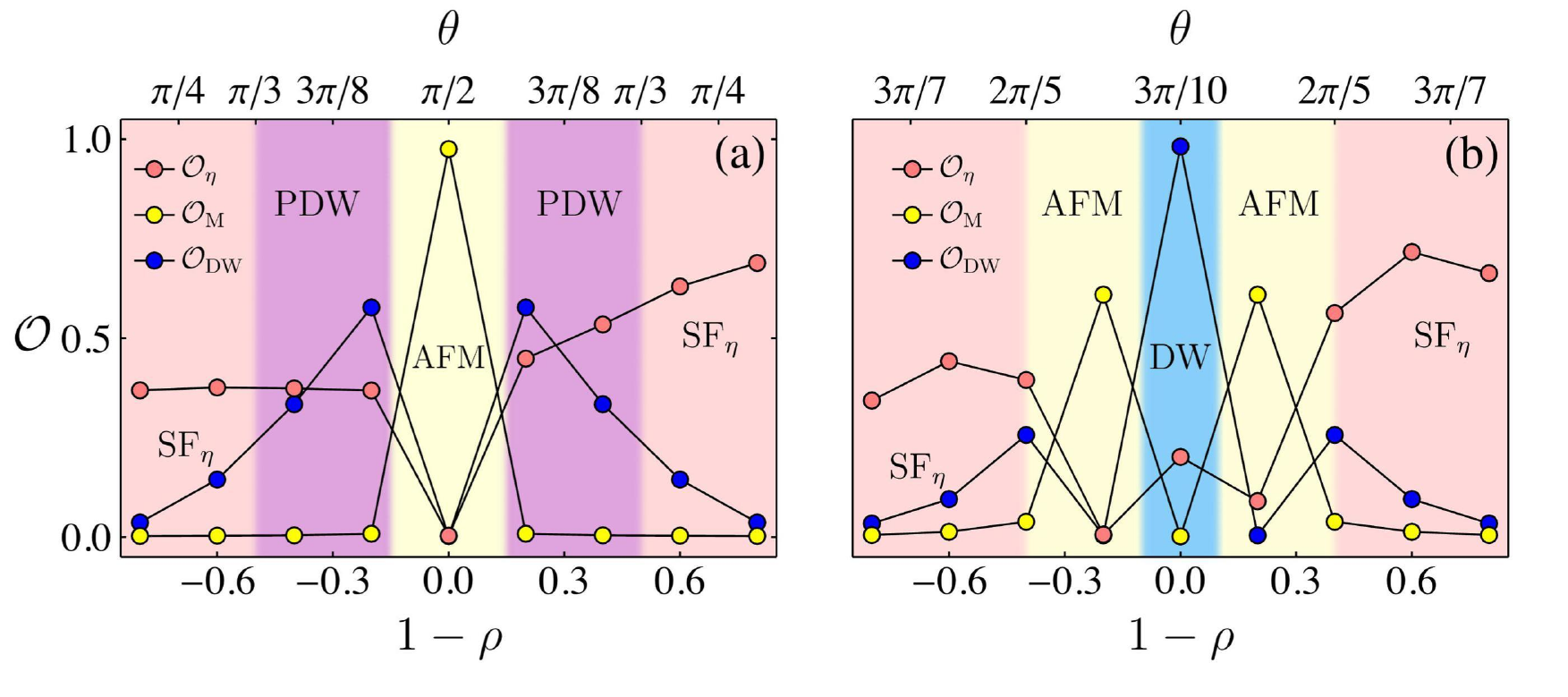} 
\end{center}
\caption{ \footnotesize (Color online) Competition of quantum phases by tuning the angle $\theta$ for given values of the doping. Order parameters $\mathcal{O}_{\textrm{DW}}$, $\mathcal{O}_{\textrm{M}}$ and $\mathcal{O}_\eta$ as  functions of the filling factor and the angle of polarization $\theta$. In panel (a), at half-filling an AFM phase is produced with $\theta=\pi/2$. Properly changing the angle of the linear polarized beam to $\theta=3\pi/8$, a PDW phase emerges doping the system by one or two holes/pairs. For values of doping far beyond of half-filling and $\theta=\pi/4$, a SF$_\eta$ phase dominates. In panel (b), a DW insulator is produced at half-filling with $\theta=3\pi/10$. Doping by one pair/hole and changing the polarization angle to $\theta=2\pi/5$, the system is an AFM.  A SF$_{\eta}$  phase can be accessed away from half-filling by doping further and increasing the polarization angle to $\theta=\pi/4$. Parameters are:  $N_s=10$, $t/|U|=0.1$, $g_{\mathrm{eff}}/|U|=-0.75$ (a) and $g_{\mathrm{eff}}/|U|=-1$ (b).}
\label{Fig4}
\end{figure}

\textit{Engineering competition of quantum phases and order parameters.}  The polarization of the pump beam controls the effective spatial and inter-species coupling in~the QOL through   $\hat{D}_\theta$, the polarization of light along $-x$ axis in~the effective Hamiltonian couples the density modes via $\hat{D}_x=\hat D_{\theta=0}=\sum_{j}(-1)^j\hat\rho_{j}$, with $\hat\rho_{j}=\hat{n}_{j\uparrow}+\hat{n}_{j\downarrow}$. The polarization in $-y$ direction couples the fermions via the staggered magnetization operator $\hat{D}_y=-i\hat D_{\theta=\pi/2}=\sum_{j}(-1)^j\hat{m}_j$, with $\hat{m}_j=\hat{n}_{j\uparrow}-\hat{n}_{j\downarrow}$, $\langle \hat m_j\rangle$ is the local magnetization. For $g_{\mathrm{eff}}<0$, the effective interaction induced by light creates a staggered field breaking the translational symmetry generating DW order predominantly for $0\leq\theta<\pi/4$, while inducing predominant antiferromagnetic order for $ \pi/4<\theta\leq\pi/2$. Indeed, at $\theta=\pi/4$, maximum competition between orders takes place, as they both become favoured in~the same proportion. The angle of the linear polarization of the pump beam, together with doping (number of holes) allow to engineer phase paths to study the competition between several quantum phases. The atoms will self-organize, optimizing the energy with maximal light scattering $g_{\mathrm{eff}}<0$ leading to superradiant states and minimal for $g_{\mathrm{eff}}>0$, enhancing quantum fluctuations\cite{Santiago3}.  In the following, we consider $g_{\mathrm{eff}}<0$ and $U<0$, on-site attraction between species for simplicity, additional results will be reported elsewhere. 
In Fig.~\eqref{Fig4} we show that different routes (trajectories) can be created. For example the route  [Fig.~\eqref{Fig4} (a)]
$$
\textrm{SF}_\eta\rightarrow\textrm{PDW}\rightarrow\textrm{AFM}\rightarrow\textrm{PDW}\rightarrow\textrm{SF}_\eta
$$ 
can be designed,
or a path [Fig.~\eqref{Fig4} (b)]
$$
\textrm{SF}_\eta\rightarrow\textrm{AFM}\rightarrow\textrm{DW}\rightarrow\textrm{AFM}\rightarrow\textrm{SF}_\eta
$$
engineered. This allows us to emulate the strongly correlated regime of fermionic quantum matter analogous to a prototypical scenario of High-$T_c$ superconductors~\cite{Fradkin}. For instance in a real system, like an electronic material, one can have competition between DW and AFM for example.  In our system the superfluid, density wave and spin order, analogous to the superconducting, charge wave and spin order parameters, show the competition between different quantum many-body phases. We show the behaviour of the order parameters associated to each phase in Fig.~\eqref{Fig4}. Those order parameters were obtained directly from the calculation of ground state correlation functions, while considering the following definitions, 
\begin{eqnarray}
S_{\textrm{M}}(q)&=&\frac{1}{N_s}\sum_{j,l}e^{iq (j-l)}(\langle \hat{m}_j\hat{m}_l\rangle-\langle \hat{m}_j\rangle \langle \hat{m}_l\rangle),\\
S_{\textrm{DW}}(q)&=&\frac{1}{N_s}\sum_{j,l}e^{iq (j-l)}(\langle \hat{\rho}_j\hat{\rho}_l\rangle-\langle \hat{\rho}_j\rangle \langle \hat{\rho}_l\rangle),\\
(\rho_{2})_{i,j}&=&\langle  \hat{f}^\dagger_{i,\uparrow}\hat{f}^\dagger_{i,\downarrow}\hat{f}_{j,\downarrow}\hat{f}_{j,\uparrow}\rangle,
\end{eqnarray}
where $S_{\textrm{M/DW}}$ is the magnetic/density structure factor and $\rho_2$ is two-body reduced density matrix.

Magnetic order is characterized through the structure factor $S_{\textrm{M}}(q)$ being $q$ the magnitude of a wavevector in~the first Brillouin zone. We use the AFM parity as order parameter $\mathcal{O}_{\textrm{M}}=S_{\textrm{M}}(q=\pi)/N_s=\langle\hat D_y^2\rangle/N_s^2$. The density wave order parameter is $\mathcal{O}_{\textrm{DW}}=S_{\textrm{DW}}(q=\pi)=\langle\hat D_x^2\rangle/N_s^2$. For two atomic spatial modes, the operator $\hat{D}_x=\sum_\sigma(\hat{N}_{e,\sigma}-\hat{N}_{o,\sigma})=\sum_\sigma\sum_{i}(-1)^i\hat n_{i,\sigma}$ where $e/o$ denotes even/odd sites. If $ \mathcal{O}_{DW}\neq 0$  the translation symmetry of the atoms in~the lattice is spontaneously broken, in 2D this leads to the characteristic checkerboard pattern in a square OL. When light scattering is different from $90^\circ$ one obtains to additional light-induced spatial modes~\cite{Santiago3}. To estimate the existence of the pair superfluid state ($\textrm{SF}_\eta$), we use the notion of off-diagonal long range order (ODLRO)~\cite{Yang}. This establishes that, given the knowledge of the two-body reduced density matrix $\rho_{2}$, if the maximum of its eigenvalues $\lambda$ scales as the system size, then the system has ODLRO. Thus, the presence of  $\textrm{SF}_\eta$ is estimated via $\mathcal{O}_\eta={4\lambda}/{N_s}$ where the factor of $4$ comes from a normalization~\cite{Yang}. Additionally off-diagonal elements in $\rho_{2}$  have to be comparable with the system size. Below and above from half-filling, we take into account that deep in the insulating DW phase, $\rho_2$ acquires non-zero off diagonal terms as a consequence of the finite size and are irrelevant for true ODLRO, such terms deep in the DW phase scale as  $\mathcal{O}_\eta|_{\textrm{DW}}\sim4/N_s^2$, vanishing in a thermodynamic limit. We compare our numerical results with this estimate to determine the possible emergence of superfluidity, provided by the criterion $\mathcal{O}_\eta>\mathcal{O}_\eta|_{\textrm{DW}}$.The ground state is found using Exact Diagonalization up to $N_s=10$ sites with periodic boundary conditions and then  operator expectation values are estimated. The use of $\mathcal{O}_\eta$ has been discussed in~the context of superconducting states in electronic systems~\cite{Essler1, Tanaka}, as well as, Fulde-Ferrell-Larkin-Ovchinnikov (FFLO) states and extended Hubbard models~\cite{Dhar}. The correspondence between the quantum phases (QP) of system and the order parameters is in~the table~\eqref{tabla1}.
 \begin{table}[!t]
\centering
\caption{Relation between  parameters and quantum many-body phases (QP)}
\label{tabla1}
\begin{tabular}{|l|l|l|l|}
\hline
\footnotesize
QP/Parameter &\footnotesize  $ \mathcal{O}_{\textrm{DW}}$ &\footnotesize   $ \mathcal{O}_{\eta}$  &\footnotesize   $ \mathcal{O}_{\textrm{M}}$ \\ \hline
SF$_{\eta}$ & 0 & $\neq 0$ & 0  \\ \hline
DW & $\neq 0$ & $ 0$ & 0 \\ \hline
AFM & 0 & 0     & $\neq 0$ \\ \hline
PDW & $\neq 0$ & $ \neq 0$     & 0   \\ \hline
\end{tabular}
\end{table}

In Fig.\ref{Fig4} we show how the formation of quantum phases can be controlled by properly orientating the linear polarizer of the pump beam, while changing doping. For each doping realization, the quantum phases can be accessed by fixing the interspecies interaction $U$, and the effective matter-light coupling strength $g_{\mathrm{eff}}$. In Fig.\ref{Fig4} (a), for $g_{\mathrm{eff}}/|U|=-0.75$, we observe AFM  emerging at half-filling by polarizing the pump beam along the $y$ axis ($\theta=\pi/2$). When doping away from half-filling, $\rho=1\pm\frac{2}{N_s}$, with the external polarizer oriented at $\theta=3\pi/8$, a PDW arises. PDW is characterized by the coexistence of DW and superfluid order. Finally, a homogenous SF$_\eta$ phase occurs when the system is doped further away from half-filling and the orientation of the polarizer is decreased to $\theta=\pi/4$.  Shaded regions in Fig.\ref{Fig4} indicate the quantum many-body phase of the atoms, as well as, the values of $\theta$ and $\rho$. As can be seen from Fig.\ref{Fig4}(b), where $g_{\mathrm{eff}}/|U|=-1$, it is possible to produce a different combination of boundaries by changing the angle of the linear polarizer.  For instance, one can generate a DW insulator by setting the polarization angle equal to $\theta=3\pi/10$ at half-filling, or induce magnetic order by simultaneously increasing the polarization angle and changing the doping of the chain ($\rho=1\pm \frac{2}{N_s}$). This AFM is achieved by selecting at the same time $\theta={2\pi}/{5}$. Further away from half-filling, the attractive on-site interaction ($U<0$), inhibits AFM order and gives rise to SF$_\eta$. Indeed, it is remarkable that engineering these quantum phases involves only the simultaneous change of doping and polarization angle of the pump beam, while all other parameters are fixed. In typical experiments with ultracold atoms, one could investigate the trajectories and the transitions between states, as well as, their reversibility~\cite{Landig}, i.e. at fixed density while changing the polarization angle. 

\begin{figure}[t!]
\includegraphics[width=0.47\textwidth]{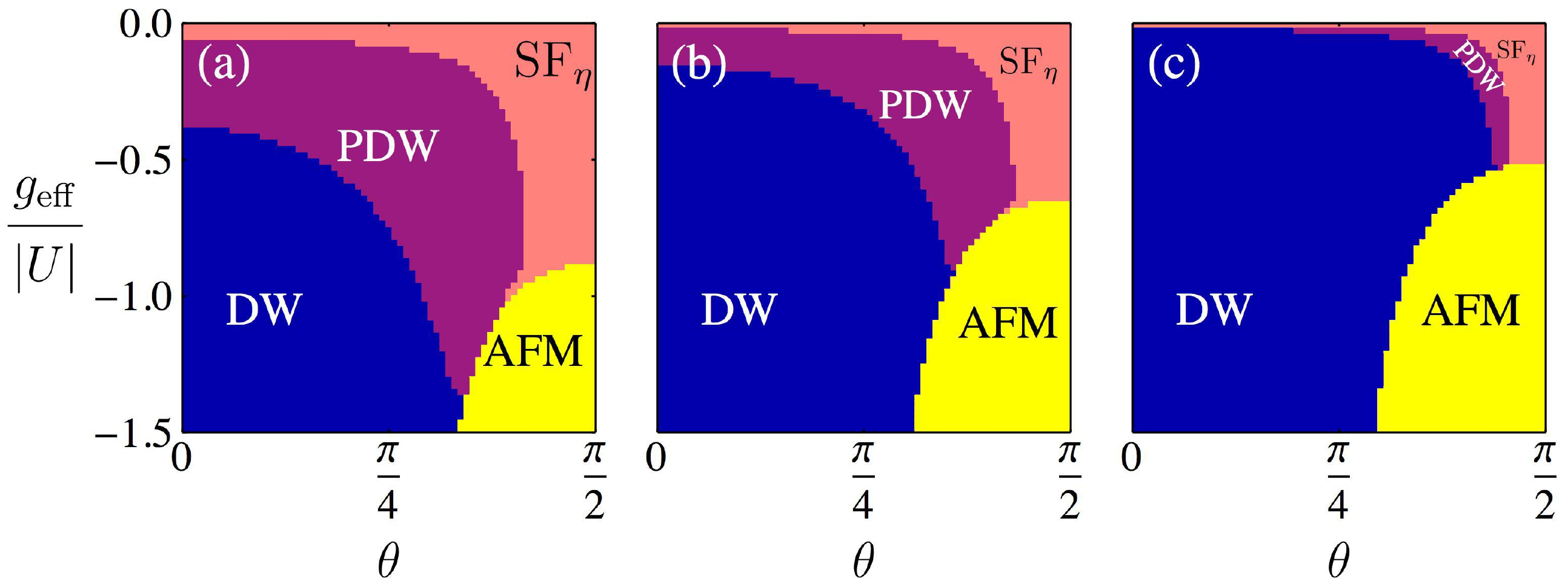} 
\caption{\footnotesize (Color online) Phase diagram of fermions in QOL with balanced population. We show competition of QP for different doping in~the strongly attractive interacting limit $|U|\gg t$. We plot the QP~in~the system for several values of doping as a function of the detuning in terms of $g_{\mathrm{eff}}$ and the angle $\theta$. Doping is: $\rho=1-\frac{4}{N_s}$ (a),  $\rho=1-\frac{2}{N_s}$ (b), and  $\rho=1$ (c). As the system is doped with holes [(c) to (a)] the regions in~the phase diagram that support a DW insulator and AFM shrink while the intermediate PDW enlarge. Depending on the value of $g_{\mathrm{eff}}$ one can access trajectories such as the ones portrayed in Fig.\ref{Fig4}. Parameters are: $N_s=10$,  $t=0.1$, $N_\uparrow=N_\downarrow$, $N_\sigma=\langle\sum_i\hat n_{i,\sigma}\rangle$. }
\label{Fig3}
\end{figure}

\textit{Phases of quantum matter and quantum domains}. Understanding the results delineated above is possible by analyzing the phase diagram of the system varying $\theta$ for $g_{\mathrm{eff}}<0$, for fixed $U$ in~the strong attractive interaction limit ($t/|U|\ll1$). For $|g_{\mathrm{eff}}|/|U|\gg 1$, the ground of the system has broken translational symmetry for $\theta<\pi/4$, generating a DW insulator, see Fig.\ref{Fig3}. On the other hand, a paired AFM phase appears for $\theta>\pi/4$. For intermediate values of $|g_{\mathrm{eff}}|/|U|$ the competition between short- and long-range interactions becomes evident. The inter-species on-site interaction favors on-site pairing (ODLRO). This competes with the long-range cavity mediated interaction favouring either DW or AFM. For finite $|g_{\mathrm{eff}}|/|U|$, the DW$\leftrightarrow$AFM transition takes place always for $\theta>\pi/4$.
When the chain is doped, see Fig.\ref{Fig3} (a) and (b), magnetic and density order remain. in~the limit where the effective light-matter interaction dominates $|g_{\mathrm{eff}}|/|U|\gg1$, DW or AFM domains surrounded by holes or double occupied sites, depending on $\theta$, are formed. The existence of holes or pairs depends on doping. For sub-doping (supra-doping) the system produces holes (doubly occupied sites) that act as borders between DW or AFM domains. For $\theta<\pi/4$ and $|g_{\mathrm{eff}}|/|U|\gg1$, the ground state is a quantum superposition of states with DW domains of the form $|0,0,|\textsc{DW}|_1,0,0,..,|\textrm{DW}|_2,0...,|\textrm{DW}|_3..0   \rangle$ with holes/pairs acting as walls. Namely, multiple domains with long-range density order result from long-range interaction instead of a single domain $|\textrm{DW}|$ of length  $\rho=1\pm\frac{n}{N_s}$. Analogously, for $\theta>\pi/4$ and $|g_{\mathrm{eff}}|/|U|\gg1$ the ground state is  a quantum superposition of states with AFM domains  $|0,0,|\textsc{AFM}|_1,0,0,..|\textrm{AFM}|_2,0...,|\textrm{AFM}|_3..0   \rangle$. The length of those domains depends on the value of the doping, tending to maximize $\mathcal{O}_{\textrm{M}}$. Surprisingly, these highly degenerate partially ordered states (AFM/DW) with domains arise and even at $T=0$, having full quantum origin. When favouring DW, away from AFM, there is a smooth transition towards SF$_\eta$ at fixed $|g_{\mathrm{eff}}|/|U|$ as $\theta$ increases.   On the contrary, the transition from a state  with  DW and/or $\eta$-pairing to the AFM is always sharp, there is no intermediate phase. This occurs as $\eta$-pairing and DW are naturally orthogonal to AFM. An intermediate PDW always occurs as the system transits to the SF$_\eta$. The coexistence region between DW and SF$_\eta$ becomes larger as we move the system away from half-filling, see Fig.\ref{Fig3}(b) for $\rho=1-\frac{2}{N_s}$. These magnetic and density modulated spatial structures, could be investigated with methods similar to those employed to study long range hidden magnetic order in doped chains using quantum microscopy~\cite{Hilker}. Moreover, in higher dimensions, we can anticipate that the competition between different kinds of domains could play a role in~the formation of resonance valence bond (RVB) states~\cite{Anderson} and other dimerized quantum matter~\cite{Balents}. 

\textit{Experimental considerations and measurement}. Recent achievements with ultracold Fermi atoms are suitable to reproduce the physics depicted in this article. It is now possible to have $^6$Li atoms in 1D lattices composed of 7-15 sites ~\cite{Boll,Hilker}, with typical tunnelling amplitudes of $t/h= 400$Hz, lattice constant $a=1.15\mu m$ and interspecies interaction energy $U/h= 2.9$ kHz. Since the Feshbach resonance of $^6$Li allows the system to have a 3D negative scattering length, an effective attractive interaction in~the chain can be achieved~\cite{Houbiers1}. Moreover, the experimental progress to have atoms with an external OL in a high-Q cavity allows for typical detunings in~the range -70MHz $\lesssim\Delta_c/2\pi\lesssim$ 20MHz with cavity decay rates of $\kappa/2\pi\sim$2MHz. Indeed, it is possible to reach the limit $\kappa\ll\Delta_c$~\cite{Landig}, with  $^{87}$Rb  atoms  (bosonic) in two dimensional OL's in a single mode cavity and an OL wavelength of $785$nm. It remains to integrate both the cavity and the ultracold fermions in the OL, which seems feasible in principle with current technology in the near future. The setup the advantage of being able to allow full external parametric control, and could be in principle be extended to allow control of other quantum objects such as molecules~\cite{Trimer}.  

The emergence of AFM and DW can be measured with single site resolution using quantum-gas microscopes~\cite{Haller, Boll,Cheuk,Mazurenko}, or alternatively, by measuring the polarization of the output photons. Since $n_{\mathrm{ph}}\sim\langle\hat D_\theta^\dagger\hat D^{\phantom{\dagger}}_\theta\rangle$, it is possible to access either order parameter $\mathcal{O}_{\textrm{M}/\textrm{DW}}$ by the proper choice of $\theta$. Similarly, DW has been measured in~\cite{Landig}. Other measurement schemes with light~\cite{Mekhov1,PRAWojciech2015} or atomic probing \cite{Elliot,Buchleitner} are possible. 

\textit{Summary and discussion}. In this article, we have shown how to control the emergence and engineer the competition of quantum many-body phases with attractive fermions inside an optical cavity illuminated by a transverse pump beam. In particular, we demonstrated that the phases that emerge DW, PDW, AFM and SF$_\eta$, depend on angle that defines the linear polarization of the pump beam while modifying value of pair/hole doping. We found highly degenerate quantum domains AFM/DW that depend on the global interaction induced by the cavity back-action and the polarization of light. We have showed that cavity backaction in~the atomic system allows to have full external parametric control of magnetic or density order emergence, just dependent on atomic loading. 

In summary, the system studied here allows to explore the competition between many-body phases of quantum matter with fully controlled mechanisms. This provides a rich experimental landscape for developing new quantum simulators possessing common features with condensed matter systems. This fosters further experimental and theoretical studies in~the higher dimensional version of our system. FQOL could be used with the aim of investigating possible analog superconducting mechanisms and their interplay with emergent orders of quantum matter. Other venues of exploration in the future include manipulation of spin or charge order~\cite{Roati,Schollock},  analog striped superfluidity control~\cite{Cavalleri}, the interplay with disorder~\cite{Lewenstein}, non-trivial band topology~\cite{TopoQu}, and exotic phases in bosonic mixtures~\cite{Soyler, Penna}.  

\textit{Acknowledgements.}
SFCB acknowledges financial support from C\'atedras CONACYT project 551. SFCB also thanks Consorcio CICESE-INAOE-CIO in PIIT, Apodaca N.L. and  IF-UNAM for their hospitality. ACG acknowledges CONACYT scholarship. This work was partially funded by grants IN111516, IN105217 DGAPA (UNAM) and 255573 CONACYT.

\end{document}